\documentclass[acmsmall,screen,nonacm]{acmart}
\AtBeginDocument{%
  }

\usepackage{subfig}
\usepackage{nameref}
\usepackage{algorithmic}
\usepackage{graphicx}
\usepackage{textcomp}
\usepackage{xcolor}
\usepackage{enumitem}
\usepackage{xcolor}
\usepackage{doi}
\usepackage{pgfplots}
\usepackage{url}
\usepackage{amsmath}
\usepackage{array}
\usepackage{tabularx}
\usepackage{pgf-pie}

\newcounter{conclusionctr}

\newcommand{\extract}[3][]{%
  \textcolor{purple}{``#2''}%
  \if\relax\detokenize{#1}\relax
  \else
    \space(P#3)%
    \fi%
}
\newcolumntype{P}[1]{>{\hspace{8pt}}p{#1}<{\hspace{0pt}}}

\newlist{questions}{enumerate}{2}
\setlist[questions,1]{label=\textbf{RQ\arabic*},ref=RQ\arabic*}
\setlist[questions,2]{label=(\alph*),ref=\thequestionsi(\alph*)}

\makeatletter
\newcommand{\conclusion}[3][]{%
  \par\refstepcounter{conclusionctr}%
  \textbf{\textit{#2:}} \textit{#3}.%
  \ifx\\#1\\%
  \else%
    \protected@edef\@currentlabelname{#2}%
    \label{#1}%
  \fi%
}
\makeatother

\begin{document}

\title{Toward Practical Deductive Verification: Insights from a Qualitative Survey in Industry and Academia}

\author{Lea Salome Brugger}
\email{leasalome.brugger@inf.ethz.ch}
\orcid{0000-0001-8770-4112}
\affiliation{%
  \institution{ETH Zurich}
  \city{Zurich}
  \country{Switzerland}
}
\author{Xavier Denis}
\email{research@xav.io}
\orcid{0000-0003-2530-8418}
\affiliation{%
  \institution{ETH Zurich}
  \city{Zurich}
  \country{Switzerland}
}
\author{Peter M\"uller}
\email{peter.mueller@inf.ethz.ch}
\orcid{0000-0001-7001-2566}
\affiliation{%
  \institution{ETH Zurich}
  \city{Zurich}
  \country{Switzerland}
}

\renewcommand{\shortauthors}{Brugger, Denis, and M\"uller}

\begin{abstract}
    Deductive verification is an effective method to ensure that a given system exposes the intended behavior. 
In spite of its proven usefulness and feasibility in selected projects, deductive verification is still not a mainstream technique.
To pave the way to widespread use, we present a study investigating the factors enabling successful applications of deductive verification and the underlying issues preventing broader adoption.
We conducted semi-structured interviews with 30 practitioners of verification from both industry and academia and systematically analyzed the collected data employing a thematic analysis approach.
Beside empirically confirming familiar challenges, e.g., the high level of expertise needed for conducting formal proofs, our data reveal several underexplored obstacles, such as proof maintenance, insufficient control over automation, and usability concerns.
We further use the results from our data analysis to extract enablers and barriers for deductive verification and formulate concrete recommendations for practitioners, tool builders, and researchers, including principles for usability, automation, and integration with existing workflows.
\end{abstract}

\maketitle

\section{Introduction}\label{sec:01}
Software verification is a cornerstone for ensuring the reliability and security of critical systems. 
Several efforts, such as the Ironclad~\cite{ironclad} project on a complete verified software stack, seL4~\cite{sel4}, a formally verified microkernel, the CakeML project~\cite{cakeml} including a verified compiler, a formally verified authorization engine developed at AWS~\cite{aws}, and the VerifiedSCION project~\cite{verified-scion} verifying a next-generation Internet router, have demonstrated that verification is useful and, in principle, applicable to large-scale projects.
Nevertheless, despite its effectiveness in selected scenarios, the practical adoption of verification in industrial settings remains limited.

While much research has focused on the theoretical foundations, techniques, and tools for verification, there has been little exploration of \emph{how} these methods should be employed in practice to maximize their benefits.
In particular, understanding the characteristics of successful applications and the difficulties faced by practitioners working on verification projects is crucial for driving the development of practically useful techniques and tools and for applying them effectively.

This paper presents the first qualitative investigation into the practical use of \emph{deductive} verification (as opposed to algorithmic verification, i.e., model checking, or lightweight bug-finding). Different verification techniques have diverse characteristics with far-reaching implications for practical applications, which necessitates a dedicated study.
Deductive verification  offers rigorous guarantees by proving that an implementation conforms to its formal specification. 
This mathematical proof can be achieved using various degrees of automation: 
Proof assistants like Rocq~\cite{coq}, Isabelle~\cite{isabelle}, or Lean~\cite{lean} support the interactive construction of machine-checked proofs, whereas
automated deductive verifiers such as Dafny~\cite{dafny}, F*~\cite{fstar}, Verus~\cite{verus}, or Viper~\cite{viper} automatically construct a proof (typically using SMT solvers) based on user-provided program annotations. 
In the remainder of this paper, ``verification'' refers to deductive verification, both interactive and automated.

The goal of our study is to answer the following research questions.

\begin{questions}[leftmargin=1.3cm,]
        \item What are the characteristics of successful application of deductive verification? \label{rq:1}
        \item What are the challenges faced by engineers when applying deductive verification? \label{rq:2}
        \item What opportunities exist to improve the usability and applicability of deductive verification techniques and tools? \label{rq:3}
\end{questions}

\ref{rq:1} aims to identify the conditions, practices, and tool features that enable verification efforts to succeed.
\ref{rq:2} addresses common pain points preventing engineers from effective and efficient application of verification in practice, both from a technical and an organizational perspective.
\ref{rq:3} aims to look beyond the status quo by providing actionable directions for enhancing verification frameworks and practices to make them more adoptable.

Through semi-structured interviews with 30 practitioners from industry and academia, we investigate these questions by discussing their experiences, difficulties, and perceptions of deductive verification and its practical application.
We evaluate the data collected from the interviews using the thematic analysis approach~\cite{thematic-analysis} and present the aggregated results. 
Crucially, we extract recommendations for practitioners, tool developers, and researchers from the findings of our analysis that are grounded in empirical data.

Besides empirically confirming familiar challenges, e.g., the high level of expertise needed for conducting formal proofs, our data analysis reveals several underexplored obstacles:
For example, maintaining proofs while software evolves emerged as a major barrier for which no mature solutions exist.
Moreover, perhaps surprisingly, automation is viewed as both a blessing and a curse because it reduces the proof effort, but also fine-grained control over the verification process.

Based on the insights from our data analysis, we formulate concrete recommendations for users and tool builders, most of which also pose research challenges.
For instance, users should focus verification on the critical components of a system; however, existing verification techniques generally do not provide strong guarantees unless the entire system is verified.
Similarly, developers should increase the usability of their frameworks, yet it is unclear how to best interface with a verification tool, e.g., when debugging verification errors.

In summary, our core contributions are the following:

\begin{enumerate}[label=(\arabic*)]
    \item We offer a detailed understanding of the interviewed practitioners’ opinions and experiences regarding practical applications of deductive verification.
    \item We distill four success factors and five main barriers for deductive verification, providing a principled account of when and how verification pays off.
    \item We highlight critical but underexplored issues such as proof maintenance and verification debugging, which require targeted attention.
    \item We translate our findings into practical recommendations for practitioners, tool builders, and researchers, including guidelines for usability, automation, and workflow integration.
\end{enumerate}

This paper is structured as follows:
Sec.~\ref{sec:02} describes our methodology and discusses threats to validity.
Sec.~\ref{sec:03} presents the results of our data analysis.
Sec.~\ref{sec:04} relates these results to our research questions, extracting success factors and obstacles, and converting them into
recommendations.

\section{Methodology}\label{sec:02}
Our study was perfromed in accordance with the ACM SIGSOFT Empirical Standard for qualitative surveys~\cite{acm-standard}.
To address the research questions listed in Sec.~\ref{sec:01}, we performed interviews with practitioners of deductive verification.
Sec.~\ref{sec:02-participants} gives an overview of the recruitment and demographics of the interviewees, whereas Sec.~\ref{sec:02-interviews} elaborates on the interviewing process.
We analyzed the collected data using the thematic analysis approach by Braun and Clarke~\cite{thematic-analysis}; details on our analysis methodology are discussed in Sec.~\ref{sec:02-data-analysis}.
In Sec.~\ref{sec:02-threads-to-validity}, we reflect on potential threats to validity.

\subsection{Participants}\label{sec:02-participants}
We mainly recruited participants via professional contacts in academia and industry and also utilized snowball sampling, with interviewees referring other potential candidates.

Interviewees were invited solely based on their experience with applications of verification, ensuring a diverse range of perspectives.
To gather some background information, all participants filled out a brief questionnaire prior to the interview, asking
participants to indicate 
\begin{enumerate}[leftmargin=1cm,label=(\roman*)]
    \item their current role, \label{qq:role}
    \item what method(s) of formal verification they use, \label{qq:methods}
    \item which tools they use, \label{qq:tools}
    \item whether they are a developer of any tool they apply, \label{qq:developer}
    \item how many years of experience they have in applying formal verification at large scale, and \label{qq:experience}
    \item how comfortable they feel applying formal verification. \label{qq:comfortable}
\end{enumerate}

The aggregated results of this pre-interview questionnaire are also available in our artifact~\cite{artifact}.

\begin{figure}[h]
    \centering
    \subfloat[Geographical distribution of the participants based on the location of their indicated affiliations' headquarters. Classification was done by the authors.]{%
        \begin{tikzpicture}[scale=0.8]
  \pie[
      radius=2,
      color = {
        yellow!90,
        cyan!90,
        pink!70}
  ]{63.3/Europe, 33.3/America, 3.3/Asia}
\end{tikzpicture}%
        \label{fig:geography}
    }
    \hfill
    \subfloat[Indicated usage of tools. Tools marked with $^\dag$ include related frameworks.]{%
        \begin{tikzpicture}[scale=0.8]
    \begin{axis}[
        xbar,
        yticklabels={$\text{Rocq}^{\dag}$, $\text{Viper}^{\dag}$, F*, Verus, Why3, Dafny, Frama-C, Lean, $\text{Isabelle}^{\dag}$, SPARK, Other},
        ytick={1, 2, 3, 4, 5, 6, 7, 8, 9, 10, 11},
        y dir=reverse,
        xmin=0, 
        xmax=11,
        xlabel={\# Participants Indicating Use},
        height=6cm,
        width=0.56\linewidth,
        nodes near coords
    ]
        \addplot[
            fill=green!40,
            draw=black
        ] coordinates {
            (8,1)
            (8,2)
            (6,3)
            (5,4)
            (4,5)
            (3,6)
            (3,7)
            (3,8)
            (2,9)
            (2,10)
            (10,11)
        };
    \end{axis}
\end{tikzpicture}%
        \label{fig:tools}
    }
    \hfill
    \subfloat[Years of experience applying verification in practice. Participants were asked to mark the most specific option.]{%
    \begin{tikzpicture}[scale=0.8]
    \begin{axis}[
        ybar,
        bar width=0.5cm,
        ylabel={\# Participants},
        xlabel={Years of Experience},
       symbolic x coords={$<$1, $<$3, $<$5, $<$10, $>$10},
        xtick=data,
        ymin=0,
        ymax=12.5,
        nodes near coords,
        width=0.48\linewidth,
        height=6cm,
        grid=both,
        grid style={dotted,gray}
    ]
        \addplot[
            fill=blue!40,
            draw=black
        ] coordinates {
            ($<$1, 1)
            ($<$3, 8)
            ($<$5, 11)
            ($<$10, 5)
            ($>$10, 3)
        };
    \end{axis}
\end{tikzpicture}%
    \label{fig:experience}
    }
    \hfill
    \subfloat[Level of comfort applying verification, indicated on a Likert scale. 1 means ``very uncomfortable'', 5 is ``very comfortable''.]{%
    \begin{tikzpicture}[scale=0.8]
    \begin{axis}[
        ybar,
        bar width=0.5cm,
        ylabel={\# Participants},
        xlabel={Level of Comfort},
        symbolic x coords={1, 2, 3, 4, 5},
        xtick=data,
        ymin=0,
        ymax=17,
        nodes near coords,
        width=0.48\linewidth,
        height=6cm,
        grid=both,
        grid style={dotted,gray}
    ]
        \addplot[
            fill=orange!50,
            draw=black
        ] coordinates {
            (1, 0)
            (2, 2)
            (3, 1)
            (4, 15)
            (5, 12)
        };
    \end{axis}
\end{tikzpicture}
    %
    \label{fig:comfort}
    }
    \caption{Results from the pre-interview questionnaire.}
    \label{fig:combined}
\end{figure}
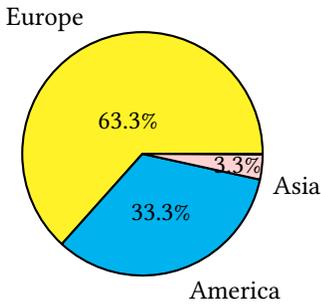
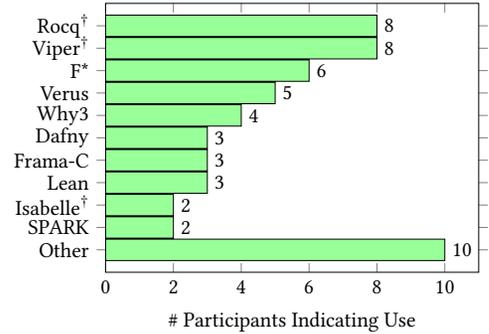
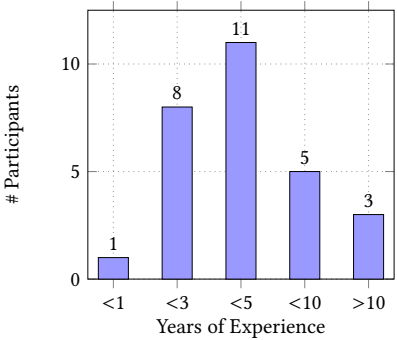
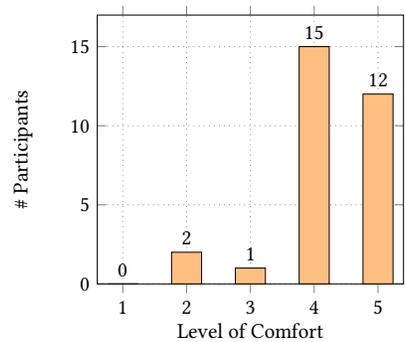

In total, we reached out to 56 practitioners and ended up with 30 study participants from 17 different institutions.
In regards to geographical distribution (see also Fig. ~\ref{fig:geography}), 19 of the participants were affiliated with an institution based in Europe, ten interviewees with an institution based in America, and one participant was at an institution based in Asia.

14 of the 30 interviewees were (mostly doctoral) students, 13 senior researchers, and three interviewees described themselves as software developers or proof engineers.
21 participants were affiliated with the academic or public sector, whereas the affiliation of the remaining nine interviewees was an industrial institution.
Despite the relatively low number of participants from industry, our study has a strong representation of industrial projects.
In 27 of the 30 interviews, we (not necessarily exclusively) discussed projects that were conducted in industrial contexts, i.e., at a company or in collaboration with an industrial partner.

Question~\ref{qq:methods} was a multiple-choice question with the options being ``automated deductive verification'', ``interactive theorem proving'', ``model checking'', and ``other (specify which)''.
26 participants indicated using automated deductive verification, whereas 15 interviewees use interactive theorem proving.
Seven interviewees said that they employ model checking or other techniques, though in all cases, they also indicated use of automated deductive verification or interactive theorem proving (or both).
17 out of 30 people stated that they are developers of (some of) the tools they use in practice.
In total, 27 different tools for deductive program verification were listed in the replies to Question~\ref{qq:tools}. 
A distribution of the indicated usage of tools is given in Fig.~\ref{fig:tools}.

The results for Questions~\ref{qq:experience} and \ref{qq:comfortable} of the survey are displayed in Fig.~\ref{fig:experience} and Fig.~\ref{fig:comfort}, respectively.
For Question~\ref{qq:experience}, two participants did not choose any of the given options, but instead stated that they do not apply verification ``at large scale'' themselves, but assist others in doing so.
Regarding Question~\ref{qq:comfortable}, interviewees were asked to indicate their level of comfort applying verification by marking the most accurate option on a five-point Likert scale.

We did not specifically aim at interviewing people until reaching saturation in our data.
Instead, we recruited a sufficiently sized group that represented the (relatively small) community of verification practitioners.
Nevertheless, as the study progressed, we observed a convergence in our results:
Later interviews did not contribute significantly to addressing our research questions compared with the already collected data, thus naturally leading to a point of saturation.

\subsection{Interviews}\label{sec:02-interviews}
The interviews were in a semi-structured format, allowing for flexibility while still ensuring that all relevant topics were covered.
Each interview was conducted via video conference, lasted approximately one hour, and was recorded and automatically transcribed using Otter.ai\footnote{\url{https://www.otter.ai}}.
Every interview was conducted by two of the authors of this study.

In preparation of the interviewing phase, we developed a guide with a set of interview questions to ask.
We performed three pilot interviews to assess the feasibility of this guide and slightly revised it before the remaining 27 interviews.
The resulting script consisted of the following questions.

\begin{enumerate}[label=(\arabic*)]
    \item Tell us about a time when you successfully applied formal verification. \label{iq:1}
    \begin{enumerate}
        \item What kinds of tools did you use?
        \item What kinds of properties did you verify?
        \item Did you verify preexisting code or implement code as you verified it?
        \item What size was the code base?
        \item How did you come up with the specifications?
        \item How did you ensure the correctness of your results?
    \end{enumerate}
    \item Tell us about a time when you dealt with a verification failure. \label{iq:2}
    \begin{enumerate}
        \item What kinds of tools did you use?
        \item What kinds of properties did you verify?
        \item Did you verify preexisting code or implement code as you verified it?
        \item How did you determine verification failed?
        \item What would you have needed to complete the proof?
        \item What did you end up doing to ensure the properties you wanted to prove?
    \end{enumerate}
    \item How is verification integrated in the workflow of software development? \label{iq:3}
    \item Which parts of verification are the most difficult or take up the most time? \label{iq:4}
    \item In which contexts is verification most useful? \label{iq:5}
    \item In which contexts is verification least useful? \label{iq:6}
    \item What are the main barriers to adopting verification more widely? \label{iq:7}
    \item Is there anything else you would like to share w.r.t.\ the practical application of verification? \label{iq:8}
\end{enumerate}

Questions \ref{iq:1}--\ref{iq:4} are targeted toward the participants' experience with concrete applications of formal methods.
The primary purpose of the subquestions in Question~\ref{iq:1} and~\ref{iq:2} was to gather all the background information necessary to put the interviewees' answers into context.
Questions \ref{iq:5}--\ref{iq:8} aim at discussing the participants' opinions on verification in practice in general.
The interview script was designed to encourage the interviewees to share their perspectives on applying verification in practice, ensuring that the conversations are general enough to retain comparability and draw meaningful conclusions.
On the other hand, the questionnaire allows for some depth by prompting participants to discuss specific real-world examples of verification projects.

While we generally adhered to the script, we deviated when relevant follow-up questions arose based on the interviewees' responses.
Such follow-up questions were primarily targeted toward clarification of or further elaboration on their statements.
Despite these deviations, we covered all scripted questions in most interviews.
In a few cases, some questions were omitted due to time constraints; however, we ensured that Question~\ref{iq:1}, \ref{iq:2}, and \ref{iq:8} were asked in every interview.

\subsection{Data Analysis}\label{sec:02-data-analysis}
For the data analysis, we followed the thematic analysis approach by Braun and Clarke~\cite{thematic-analysis}.
Thematic analysis is a qualitative method for identifying, analyzing, and reporting patterns (\emph{themes}) within data.
Such themes reflect a patterned response or meaning within the data set and highlight aspects of the data that are significant to the research questions.
Braun and Clarke divide the data analysis into six phases: familiarizing with data, generating codes (i.e., labeling segments with descriptive codes representing essential characteristics of the data), searching for themes, reviewing themes, defining and naming themes, and producing the report.

After completion of all interviews, the automatically generated interview transcripts were reviewed, corrected where necessary, and anonymized.
Due to lack of experience with deductive verification, one of the 30 interviews was excluded from the data analysis since the content of the interview was not relevant for our study.
The resulting dataset was coded in an initial coding phase by one of the interviewers using open coding, a coding method that does not assume a preexisting codebook, allowing the codes to emerge organically from the data.
The second phase of the analysis consisted of relating different codes to each other and combining them to form potential overarching themes.
Thirdly, the second interviewer reviewed the identified candidate themes with respect to coherence, lack of overlap between different themes, and ability to reflect the findings evident in the dataset.
The themes were then revised, named, and finalized in close collaboration.
The full codebook is available in our artifact~\cite{artifact}.
We present the results in Sec.~\ref{sec:03}.

\subsection{Threats to Validity}\label{sec:02-threads-to-validity}
One threat to the validity of our study arises from the composition of interview participants.
Though we actively made efforts to include a diverse range of interviewees by recruiting individuals from both academic and industrial institutions, many of the participants were affiliated with academia rather than industry.
As a result, the perspectives we gathered may not fully represent the broader verification community.
Academic participants are likely to have different incentives, priorities, and experiences with verification tools compared to their counterparts in industry, which could skew the findings toward a more research-oriented view.
However, this threat is mitigated by the fact that the majority of the interviews centered on applications of verification in industrial contexts.

The majority of our participants are researchers, many of them even developers of the tools they indicated to use, who have a vested interest in verification. 
This background may have contributed to a more positive outlook on the utility and future potential of deductive verification tools. 
The participants’ familiarity with formal methods and their academic exposure to these tools may result in an overestimation of their practical adoption and effectiveness in real-world settings.
During the interviews, we asked participants to reflect critically on both the strengths and limitations of the verification tools they use, rather than focusing only on the positive aspects. 
This helped reduce the potential for overestimating the usefulness of verification.

Another potential source of bias stems from the design of the study itself, as the data are collected from interviews, with the interviewees self-reporting on their experiences.
Participants may have difficulty accurately recalling past events, or they may unintentionally distort their memories to fit a preferred narrative, leading to unreliable data.
To mitigate this threat, we encouraged accurate recall by asking participants to describe specific instances or events and consciously avoided asking leading follow-up questions during the interviews.

Availability bias also threatens the validity of our results:
This bias describes the tendency to rely disproportionately on the most readily available data when forming opinions, thus basing them on especially memorable or recent examples and generalizing from those, instead of considering the entirety of one's experience.
Though availability bias can never be fully eliminated, we counteracted its potential impact on our results by systematically asking our interviewees about both successes and failures, not just memorable experiences, and not restricting the interview content to a certain number of projects, time frame, or domain.

Lastly, the study was conducted by researchers with expertise in verification, which introduces the risk of confirmation bias. 
Our familiarity with the field may have influenced how we framed questions and interpreted responses, potentially favoring insights that align with our prior understanding of the topic. 
Despite efforts to ask open-ended questions and to remain neutral in the interviews, our interpretations of the data could be unintentionally shaped by our own perspectives.

\section{Results}\label{sec:03}
In this section, we summarize the results of our data analysis.
We extracted five overarching themes from the data resulting from our interviews.
Each of the following subsections discusses one of them.
To preserve anonymity, we do not quote interviewees by name, but refer to them as ``P'' followed by a number for unique identification.
Unless stated otherwise, statements apply to both interactive and automated verification.

In general, the most common application domains of the projects discussed in the interviews were algorithms and data structures, cryptography, security, and systems code.
In terms of verified properties, safety and correctness were dominant, followed by security and input/output properties.

\subsection{Verification Applicability}
We outline how practitioners rationalize the investment in verification, discussing its advantages, patterns to evaluate the cost-benefit ratio, and criteria that aid in deciding when to apply it.

\paragraph{Benefits of Verification} Beyond the formal guarantees offered by verification, the most significant indicator of success in verification projects is uncovering errors.
Many interviewees (12/29) described a concrete situation where they discovered bugs in a codebase or its specifications as a successful application of verification.
In particular, interviewees (4/29) found verification useful in detecting common mistakes and helpful \extract[s]{to avoid stupid things}{14}, especially in early phases of development.
Positive side effects of verification include a better understanding of the implementation (5/29), which can even ultimately lead to \extract[s]{unlocking faster or more optimized code}{29}, and improving code quality (2/29).

\paragraph{Cost-Benefit Assessment Patterns}
Various participants (10/29) brought up cost-benefit analysis in the context of adoption barriers, stating that verification is \extract[s]{ridiculously expensive}{05} or \extract[s]{mostly doesn't pass the benefit}{02}.
P08 said that \extract{today we are just not in a place where for most companies, it makes sense [to apply verification], except for specific instances.}{08}
The criticality of software is the most significant factor in balancing investment and payoff; most interviewees (22/29) cited criticality as a key criterion for deciding when verification is useful.
Another aspect to consider is the level of adoption of the code (8/29): \extract[s]{Code that's used by many people probably benefits from some formal analysis}{09}.
Persuading non-experts in verification of the technique's value can be a challenge in itself, especially in industry (6/29); as P14 says, \extract{we need to give better arguments that [...] proof[s] are better than tests.}{14}

\paragraph{Domain Selection Criteria}
A key factor for the applicability of verification is the well-definedness of the requirements.
Eleven out of 29 participants stated that verification is feasible only when there are well-defined requirements, or, on the contrary, of limited use when requirements are unclear.
This is mostly due to the formulation of specifications becoming intractable when the software's expected attributes are ambiguous or nebulous.
Additionally, proof maintainability plays a crucial role (4/29); verification is more applicable in contexts where code does not evolve much (or, in the words of P17, \extract{[i]t's good if [software is] stable, more or less, so you don't have to completely modify it on a daily basis}{17}) and the verified software is \extract[s]{self-contained}{23}, like, e.g., library code.
Other considerations include realizability (9/29) -- whether verification can realistically be achieved -- and the size and complexity of the system (6/29).

\subsection{Managing Technical Complexity}\label{sec:03-technical-complexity-management}
This theme outlines the core technical barriers of applying verification, independent of tool specifics.

\paragraph{Fundamental Challenges}
Unsurprisingly, the interviews revealed various fundamental challenges that complicate verification.
These include the use of quantifiers and triggering (8/29), dealing with induction and recursion (5/29), and loop invariants (2/29).
In terms of properties, liveness (3/29) and global properties (2/29) were noted as especially challenging.

\paragraph{Proof Strategy}
Twelve out of 29 participants identified the process of doing the actual proof as one of the most challenging parts of verification; these include users of both automated tools (9/12) and proof assistants (4/12).
This process includes coming up with a strategy to structure a proof, iterating between specification, program annotations, and proof, or as P09 put it, \extract{getting the computer to agree with you.}{09}
The complexity of verifying code partly stems from the amount of details that have to be considered in verification, which four participants pointed out.

One question that came up in six interviews was whether to choose a top-down (starting at the root of the method call graph) or bottom-up (starting at the leaves) approach in verification.
There was no clear conclusion in the data; which one is better suited depends on the verified properties, the scope of the project, and personal preference.

When it comes to structuring verification proofs, scalability plays a tremendous role.
Many participants (12/29) highlighted the value of compositional reasoning and modular proofs in mitigating performance issues when using automated verifiers.
The possibility to verify a single method or to isolate a single code block within a method were mentioned as useful features in existing verification frameworks.
However, practitioners do not always feel sufficiently supported by their tools; P14 stated that splitting up a proof into different sub-proofs \extract{is very manual}{14} and it requires expertise \extract{to know where to cut properly the analysis into subanalyses.}{14}

On a more technical level, several participants mentioned that reducing the proof context for SMT-based tools aids in overcoming performance issues or proof instability (8/29).
In practice, this mostly means \extract[s]{rephras[ing] things to be more friendly to the verifier}{09}, which, on the other hand, reduces the level of automation.
For example, context reduction can be achieved by hiding implementation details from the proof scope or by extracting intermediate lemmas (6/29).
An alternative feature or technique is to check complex proof obligations in isolated queries (2/29).

Abstraction was brought up in many interviews (7/29), most of them about automated tools, to improve scalability (as P27 put it, \extract{if you have to go for a coffee every time you verify a method, that's a good indication to start abstracting}{27}) or to facilitate proof management.
Finding the right level of abstraction introduces new challenges and was generally described as a complex task (6/29), e.g., by P08 who stated, \extract{you [...] have to guess what you will need and get that into a sufficiently precise shape that you can validate whether this works together with the rest of the system.}{08}

Problem simplification is another strategy, applied by users of both interactive theorem provers and automated verifiers, to deal with performance issues and verification failures (5/29).
This could mean \extract[s]{reducing our expectations}{13}, i.e., removing requirements from the specifications.
In the context of concrete verification errors, it was found to be useful \extract[]{to simplify the code as much as possible, and then try to find a version that works, and then work your way back up to the point what the actual code was}{24}~(P24).
Three interviewees even mentioned having constructed a simplified model of the original system and verified this model, to make verification more manageable.

A concrete instantiation of both abstraction and simplification can be found in the extensive use of refinement as a proof structuring technique (12/29). 
Certain participants went so far as to use chains of successive refinements to move from a mathematical description down to an executable program.
Refinements can serve as a hedge against unexpected proof complexity: \extract[s]{It's mainly when we're nervous and uncertain about how hard it's going to be to match the implementation to the specification that we start putting in these layers}{28}. 

\paragraph{Specification Engineering}
Numerous interviewees said that writing specifications is hard, especially in the context of correctness properties, for several reasons.
First, the sheer amount of specifications to write can be overwhelming (4/29).
Second, stating clear and accurate specifications is crucial, but also challenging (4/29).
As P22 said: \extract{Transcribing the user intent is [...] the bottleneck. [...] What do you want this piece of code to do? You had an intent, but getting that out of the head and onto some [...] natural language or formal form is [...] the bottleneck.}{22}
This issue is aggravated by the difficulty of expressing requirements in a formal way (7/29), since \extract[s]{not everything is easy to write formally, or even possible to write formally}{23}.
Third, finding the right strength of a specification poses another challenge (5/29), or, in the words of P08: \extract{finding the right point in the tradeoff of specifying something that is useful and strong enough that it's worthwhile to prove that, but also where it's feasible to prove.}{08}

Even with a positive verification result, \extract[s]{you're not quite done yet, because you have to know that your specification is good}{01}.
Participants discuss several strategies to assess the correctness and accuracy of specifications, such as, for instance, unguided manual review, which is challenging since \extract[s]{specifications are harder to understand than the code}{27} and error-prone.
Other examples are checking specifications against existing documentation or concrete test cases.

\subsection{Tool Maturity}
Remarks on complications emerging from tool (im)maturity led to four key subthemes.

\paragraph{Automation Gaps}
Automation or lack thereof was a common subject evoked by participants, with 10 out of 29 stating that more automation would be useful or that lack of automation was a challenge.
This issue affects users of both proof assistants and automated verifiers.
The former were generally described as requiring \extract[s]{too much work}{17} due to fundamental lack of automation, and features of interactive tools that allow for automating parts of the proof were praised.

Users of automated frameworks appreciated the general usefulness of sparing them from having to flesh out proofs by hand, but expressed demand for even more automation, especially of simple tasks.
For example, \extract[s]{it would be very great to not have to write [...] specification that could be inferred automatically}{06}.
LLM and AI technologies were looked forward to as a possible avenue for specification and proof generation in both interactive as well as automated tools (8/29), stating that AI \extract[s]{has the chance to change the game quite a lot, [...] change the cost-benefit landscape}{29}.

According to P27, another pain point with SMT-based tools is managing the proof context:
\extract[]{If you have more context lying around, then [the SMT solver] [...] will choke, so you hide everything.}{27}
Controlling \textit{when} to hand \textit{what} information to the SMT solver \extract[s]{makes your life more difficult, because you have to do manually part of the proof}{27}.
Overall, seven users of automated tools describe the necessity of ``guiding'' automation, e.g., \extract[s]{you [...] need to be extra careful about controlling the context so that the SMT solver does not saturate completely, and when it saturates, you [...] don't even manage to prove anymore [...] one plus one equals two}{17}.

\paragraph{Performance Limitations}
Performance was an omnipresent topic among practitioners working with automated verifiers, primarily as a source of concern (16/29). 
The threshold for slow proofs varied by participant and tool, e.g., \extract[s]{if it takes more than ten seconds, I get annoyed}{02} or \extract[s]{if your verifier cannot give you [...] meaningful feedback under [...] 30 seconds, that is unbearable}{21}.
Nonlinear arithmetic was noted (4/29) as a specific cause of performance problems. 

\paragraph{Tool Ecosystem}
The lack of sophistication of verification tools was described as a source of difficulty in various manners. 
Missing support for features of the analyzed language forced participants to change their code for verification~(9/29). 
The concrete problematic features varied by tool and included floating points, generics, and support for aliasable pointers.
Five participants lamented the lack of documentation of tools or emphasized its usefulness, and three interviewees noted the helpfulness of user support, e.g., by tool developers.

\paragraph{User Experience}
The user experience of verifiers surfaced in many different aspects. 
The general lack of maturity in tools (13/29) was common refrain, as \extract[s]{there hasn't been [...] investment in these really usable tools, meaning ones that anybody can download and run in their code and get something out of}{01}.
This is highlighted as a specific risk for industrial applications by P30 who said that \extract{if the tooling is not mature enough, [...] people [...] will try out [...] this new way of working, and they run into a barrier [...] and they go to their familiar way of working.}{30}
Eight participants, primarily ones using automated tools, pointed out the importance of a good user interface and elaborate tooling; as P30 said, \extract{if you want to introduce formal techniques [...], the tooling has to be really, really good.}{30}
Users would feel supported, for instance, by \extract[s]{having an integrated IDE that shows you both the spec[ification] and the code and lets you refactor both at the same time}{18}.
Other examples for helpful features that were mentioned include linting, syntactic coloring, package managers, and autocompletion.

Fine-grained control over the verification process was described as desirable (5/29), including debugger-like functionalities (4/29). 
Identification (8/29), localization (4/29), and presentation (4/29) of errors were mentioned as specific sources of difficulties.
As P02 expresses, \extract{it's a common source of frustration that things don't work, and it might be the tool; [...] it might be the SMT solver is misbehaving; it might be, [...] the software you're trying to verify is buggy; and it's hard to [...] distinguish between these cases.}{02}
Axiom profilers (2/29) or the ability to produce counterexamples in the case of failed verification attempts (10/29) were highlighted as useful features for debugging.

Lastly, automated verification tools are often seen as black boxes (9/29) and access to more information about the internal state of the tool was asked by many participants (8/29).
As P07 stated, \extract{many of us have felt the pain of the SMT [solver] just saying, sorry, I don't know, [...] and it's always frustrating, and you have to kind of fight with it}{07}.
Features giving the user insight into \extract[s]{which proof obligation [the tool] is actually doing}{09} or \extract[s]{what quantifiers are being instantiated}{20} were regarded as essential.
Moreover, participants expressed a desire to view intermediate states, e.g., by inspecting \extract[s]{some kind of list of facts that I know up to [a certain] point}{15} or \extract[s]{some map where we can look up at a specific point what permissions do you hold}{24}.

\subsection{Integration of Verification and Development}
Another key theme concerns aligning verification with development workflows.

\paragraph{Integrating Verification}
Verification \emph{tools} must relate closely to the technologies already in use, or have a low entry barrier.
As P16 stated, \extract{you cannot ask engineers to change completely [...] the way they work, to switch to something different.}{16}

To ease the integration of verified \emph{code}, four interviewees discussed verifying code in a dedicated language and extracting source code from it. 
Interfacing with unverified components poses a major challenge (8/29):
Often, verification relies on assumptions about how a verified method is called, which are not guaranteed to hold in an unverified context.
Similarly, invoking unverified methods, e.g., library code, in a verified method may require specifying contracts without verifying them, introducing unchecked assumptions.
P13 summarized this point: \extract{You have your verified code, then you have the real world where [...] your code should run, and you have to bridge the gap.}{13}

The last axis of integration concerns the human aspect, in particular, the collaboration across different teams.
Many participants found the communication with non-experts in verification challenging (15/29), with one of them concluding that \extract[s]{you either say this [...] works and you have to trust me, or you [...] lay out all the issues}{13}.
One common strategy to facilitate the communication of verification results to non-experts is to use concrete (counter)examples.

\paragraph{Coexistence of Code and Proof}
Various participants stressed the importance to consider code readability (3/29).
This is especially prevalent when working with tools generating correct-by-construction code.
P28, for example, reported: \extract{A lot of the engineering teams really pushed back on this automatically generated code and said, `We would not check this code into our repository. It looks horrible. No human would write this.'\thinspace}{28}
Readability is also a concern with automated frameworks, which require code to be interleaved with annotations. These \extract[s]{are polluting [the engineers'] nice piece of code}{23}.

\paragraph{Proof Maintenance}
Proof maintenance is an open issue that a lot of interviewees addressed (12/29).
Verification introduces an additional layer to the already complex question of how to maintain software: \extract[s]{Any large project is hard to maintain, and especially if [...] the code needs to change and [...] kept in sync with proofs and specifications}{08}. 
On a related note, many participants (8/29) lamented proof instability, with one of them saying, \extract[s]{you slightly tweak a definition, then you have random proofs breaking CI, this is really annoying}{17}.
Strategies to avoid or mitigate maintenance problems are taking immediate action once a proof fails (1/29) and considering and checking for maintainability of proofs while working on them (4/29).
Regarding the latter, concrete practices seem to be mostly tool- or project-specific, though P17 advised to \extract{make the proofs as small as possible}{17} in general, a recommendation which complements the arguments for modularity and compositionality presented in Sec.~\ref{sec:03-technical-complexity-management}.

\paragraph{Project Management}
Regarding project management, three participants emphasized the importance of integrating verification early in the development process, for two main reasons:
First, employing verification \extract[s]{earlier on in the development lifecycle finds bugs earlier}{01}.
Second, parallelizing code development and verification or writing the code with verification in mind is beneficial because \extract[s]{how the code is structured helps immensely to verify it}{05}.
This observation is in line with reports on rewriting code in order to make verification succeed (5/29).
In terms of concrete actions in ongoing verification efforts, many interviewees (11/29) reported having used CI and code reviews (3/29) in their projects and finding them very useful for regression testing and to ensure reproducibility of results and maintainability.

Most importantly, though, 13 interviewees advocated for the combination of verification with other methods for quality assurance, e.g., testing or static analysis, or reported on combining verification with such techniques in their projects in order to complement verification and compensate for its limitations.
As P29 put it, using deductive verification \extract{for the entire code base is just not tractable as of [...] today, so [...] we need more of these combination works.}{29}

\subsection{Verification Expertise}
Next, we explore how the level and nature of required expertise shape the verification process.

\paragraph{Learning and Training}
19 out of 29 participants said that working on verification requires a lot of expertise, hindering adoption.
P03, for example, stated that \extract{compared to writing software, verification is [...] many orders of magnitude more difficult.}{03}
Nine participants expand on the comparison of programming and verification, saying that regular software engineering and verification engineering demand for different ways of thinking; as P09 put it, \extract{there's a mismatch between the software engineering [...] and the verification engineering mindset.}{09}
Beyond the need for cultivating a verification mindset, many participants (12/29) highlighted the steep learning curve associated with verification tools (both interactive and automated), further complicating adoption.

\paragraph{Knowledge Transfer}
Sharing knowledge about proofs is challenging according to several participants.
This aspect encompasses the documentation of verification techniques themselves (e.g., \extract[s]{a good database of proven techniques}{09}), which two participants explicitly wished for.
Various interviewees (6/29) also emphasized the necessity to document and publish proofs and their assumptions; in P07's words, \extract{being very clear [...] about the theorems and what the guarantees are, that's something that is really needed if we want more people to take this seriously and succeed.}{07}

\section{Discussion}\label{sec:04}
This section delves into the interpretation of the findings from Sec.~\ref{sec:03}, connecting them to the research questions from Sec.~\ref{sec:01}.
We first identify the key factors contributing to success in verification projects and the most pressing challenges, and then derive opportunities for future research and development efforts grounded in the interview data.

\subsection{Status Quo (RQ1, RQ2)}\label{sec:04-status-quo}
We address \ref{rq:1} and \ref{rq:2} by discussing key results from Sec.~\ref{sec:03}, summarized in Table~\ref{tab:observations}.

\begin{table}[!ht]
    \centering
    \begin{tabularx}{\textwidth}{c|X}
    \textbf{ID} & \centering \textbf{Observation}  \tabularnewline
\hline
   \nameref{concl:sf1} & Verification efforts should focus on the most critical system components. \\ \hline
    \nameref{concl:sf2} & Introducing verification early in the development workflow is beneficial. \\ \hline
    \nameref{concl:sf3} & Tool maturity is a key enabler for adoption. \\ \hline
    \nameref{concl:sf4} & Verification has to be easily integrable into existing workflows. \\ \hline
    \nameref{concl:ba1} & Verification often does not pass the cost-benefit analysis. \\ \hline
    \nameref{concl:ba2} & Specification and annotation effort drastically increases complexity. \\ \hline
    \nameref{concl:ba3} & Verification requires a lot of expertise and a distinct mindset. \\ \hline
    \nameref{concl:ba4} & Automation is both a blessing and a curse. \\ \hline
    \nameref{concl:ba5} & Maintaining proofs as the code evolves greatly constrains applicability.
\end{tabularx}
    \caption{Extracted success factors (SF) and barriers (BA).}
    \label{tab:observations}
\end{table}

\subsubsection{Success Factors}\label{sec:04-success}
We extract four main success enablers in verification projects from the data.
\conclusion[concl:sf1]{SF1}{Verification efforts should concentrate on the most critical system components}
There is broad consensus among the interviewees that targeting critical or highly used components that demand for strong guarantees aids in ensuring that verification passes the cost-benefit analysis (see also \nameref{concl:ba1} below).
Complementing deductive verification with other techniques for less critical code helps compensate for its limitations, enhancing trust in the overall system while keeping quality assurance efforts manageable.
In several projects discussed in our interviews, verification has been successfully combined with more lightweight methods, with the general tenor being that this hybrid approach improves efficiency.

\conclusion[concl:sf2]{SF2}{Introducing verification early in the development workflow is beneficial}
Practitioners agree that integrating verification early in the development workflow is crucial to align the verification process and the overall system design.
By designing programs to facilitate their verification, verification effort can be significantly decreased.
Nonetheless, intertwining code development and the process of verifying it can be a double-edged sword.
Our interviewees feel that code that evolves a lot is harder to verify due to increased proof maintenance (see also \nameref{concl:ba5}):
Changes to the codebase can invalidate existing proofs, thereby increasing the verification overhead.
Thus, we conclude that while early and continuous verification is in general desirable, it is advisable to balance it with stability in design to avoid a continual cycle of re-verification.

\conclusion[concl:sf3]{SF3}{Tool maturity is a key enabler for adoption}
Good tooling, user-friendly interfaces, and a rich tool ecosystem were described as vital for promoting accessibility and improving the efficiency of verification tools.
Automation further promotes these benefits, provided that users have the option to intervene manually where necessary.
Although a trade-off appears to exist between the degree of automation and user control as well as tool performance (see also \nameref{concl:ba4}), participants generally express a favorable view of automated solutions.

\conclusion[concl:sf4]{SF4}{Verification has to be easily integrable into existing workflows}
The integrability of verification and corresponding frameworks has a major influence on whether it is or can be employed in industrial practice.
This issue encompasses various aspects:
First, it must be ensured that verification can be aligned smoothly with the development practices prevailing in a project, such as choice of programming languages and IDEs as well as CI/CD workflows; on the flip side, requiring engineers to adapt to completely new workflows or confronting them with unfamiliar user interfaces on top of a different way of thinking about code (see also \nameref{concl:ba3}) drastically complicates the integration of verification.
Second, concerns as simple as code readability are often overlooked by tool developers in favor of extending and refining tool capabilities, although they play a vital role in deciding whether and how verified code is integrated in a software project, as became apparent in our interviews.
Lastly, measures that aid in explaining verification results to non-experts like the ability to produce counterexamples in the case of failed proofs facilitate communication between different stakeholders, support arguments for the approach's usefulness, and improve accessibility, thus boosting efficiency and generally promoting adoption.

\subsubsection{Barriers}\label{sec:04-barriers}
We identify five major obstacles in the data for widespread adoption of verification.

\conclusion[concl:ba1]{BA1}{Verification often does not pass the cost-benefit analysis}
While our interviewees generally value verification for its formal guarantees, not all of them are convinced of its broad applicability.
The substantial cost of applying verification, a difficult and time-consuming task (see also \nameref{concl:ba2}) that demands for highly specialized knowledge (see also \nameref{concl:ba3}), was universally highlighted as major downside.
This is especially problematic in industrial contexts, where financial concerns and time constraints have to be considered, leading practitioners to see this technique as worthwhile only in specific contexts (see also \nameref{concl:sf1}).

\conclusion[concl:ba2]{BA2}{Specification and annotation effort drastically increases complexity}
The necessity to augment the code with annotations was viewed as critical factor causing verification to be complex:
Providing annotations adds additional work for developers, requiring careful attention to detail and a thorough understanding of high-level proof mechanisms and sometimes even specifics about the tool in usage. 
Balancing the effort to write and maintain these annotations with the actual development process can be a considerable burden.

Similarly, formalizing requirements as specifications is seen as a daunting task.
Stating clear and unambiguous specifications in a formal way is crucial for successful verification, yet developing and maintaining these specifications and in particular, finding the right level of abstraction, is difficult.
Surprisingly, our interviews revealed that in some cases, specification engineering is the bottleneck when employing verification, as opposed to the well-known issue of completing proofs:
Translating user intent into formal specifications poses a substantial challenge in practical applications and introduces an unexpectedly early point of failure.

\conclusion[concl:ba3]{BA3}{Verification requires a lot of expertise and a distinct mindset}
The know-how needed to conduct proofs is one of the major obstacles to a broader adoption of verification. 
Verification requires specialized knowledge, and the learning curve associated with acquiring the skills necessary for formal proofs and using verification tools significantly impedes practical applicability.

Further, our interviewees stressed that verification demands for a different mindset than regular software engineering, indicating an unexpected cognitive barrier.
Conducting formal proofs forces engineers to think differently about code and its correctness, which comprises an issue with high practical impact that is often not addressed in the literature or software engineering education.

\conclusion[concl:ba4]{BA4}{Automation is both a blessing and a curse}
Despite the value of automation being common refrain (as described in \nameref{concl:sf3}), complaints about opaqueness of existing automated frameworks and lack of user control imply shortcomings of verification tools with considerable and often neglected practical impact:
In the face of inevitable limitations to automation, balancing automated proof construction with support for manual intervention is crucial. 
Practitioners wish for more fine-grained control and more efficient and effective techniques allowing for guidance of automated verification tools, which raises issues that have clearly not been sufficiently addressed by the verification community.

In addition to lack of control over the verification process, practitioners observed that automated tools can be (much) slower than they wish or require them to be, making it difficult to iterate and to integrate verification smoothly into fast-paced development workflows.
To counteract scalability limitations, several strategies were mentioned, among which the effectiveness of compositional reasoning, modularity, and abstraction was consistently emphasized.
However, these mitigation strategies result in increased complexity and additional manual effort as most existing techniques boil down to controlling the proof context by introducing additional annotations.

\conclusion[concl:ba5]{BA5}{Maintaining proofs as the code evolves greatly constrains applicability}
Proof maintenance was identified as a pain point in practical applications by numerous interviewees, the extent of which even leads to the consensus that verification efforts should target ``frozen'' code rather than evolving software.
Given the inevitability of code evolution, this recommendation significantly impacts the feasibility of verification in practice.
Moreover, the preference for stable codebases contradicts the idea that verification should be integrated early during development (see \nameref{concl:sf2}).

Consequently, the unavailability of effective methods to adapt proofs to changes in a program not only dramatically impairs verification applicability, but also hinders optimal integration.
While the difficulty of proof construction is well-established, this nuanced aspect is often neglected.

\subsection{Outlook (RQ3)}
We now address the future-facing \ref{rq:3} by presenting our synthesized suggestions in response to the observations outlined in Sec.~\ref{sec:04-status-quo}.
We distinguish between recommendations for engineers (considering) employing verification and proposals for verification tool developers and the overall verification research community.
Table~\ref{tab:recommendations} provides an overview.

\begin{table}[!ht]
    \centering
    \begin{tabularx}{\textwidth}{c|X|c|>{\centering\arraybackslash}p{1.8cm}}
    \textbf{ID} & \centering \textbf{Recommendation} & \textbf{Addresses} & \textbf{Targets} \\
\hline
    \nameref{concl:ru1} & Consider verification in each phase of software development, especially in early stages. & \nameref{concl:sf2} & managers \\ \hline
    \nameref{concl:ru2} & Concentrate verification efforts on critical components. & \nameref{concl:sf1}, \nameref{concl:ba1} & managers \\ \hline
    \nameref{concl:ru3} & Invest in training and education. & \nameref{concl:ba3} & managers, developers \\ \hline
    \nameref{concl:rd1} & Develop methods integrating verification with other quality assurance techniques. & \nameref{concl:sf1}, \nameref{concl:ba1} & researchers, tool builders \\ \hline
    \nameref{concl:rd2} & Facilitate the management of specifications and annotations. & \nameref{concl:ba2}, \nameref{concl:ba5} & researchers, tool builders \\ \hline
    \nameref{concl:rd3} & Invest in tool usability and integrability. & \nameref{concl:sf3}, \nameref{concl:sf4} & tool builders \\ \hline
    \nameref{concl:rd4} & Balance automation with interactiveness. & \nameref{concl:sf3}, \nameref{concl:ba4} & researchers, tool builders \\ \hline
    \nameref{concl:rd5} & Improve the performance of automated tools. & \nameref{concl:sf3}, \nameref{concl:ba4} & researchers, tool builders \\ \hline
    \nameref{concl:rd6} & Facilitate proof maintenance. & \nameref{concl:ba5} & researchers, tool builders \\ \hline
    \nameref{concl:rd7} & Investigate the prevalent approach to teaching verification. & \nameref{concl:ba3} & educators
\end{tabularx}
    \caption{Recommendations for users (RU) and tool developers or researchers (RD).}
    \label{tab:recommendations}
\end{table}
\newpage
\subsubsection{Recommendations for (Prospective) Users}\label{sec:04-recommendations-users}
We offer three proposals for verification users.

\conclusion[concl:ru1]{RU1}{Consider verification in each phase of software development, especially in earlier stages}
To maximize the benefits of formal verification, we recommend integrating it early in the development process rather than treating it as an afterthought. 
Incremental verification, where proofs and specifications evolve alongside development, can help alleviate the burden of retrofitting verification onto an existing system.
Additionally, early integration promotes a mindset shift, encouraging teams to think more formally about requirements and correctness from the outset, which can lead to easier integration and collaboration with verification engineers.

These benefits could be achieved, for example, by a close collaboration between software engineers, verification engineers, and testers during the development phase, e.g., to agree on a language (subset), program design, coding practices, concurrency structure, or the interplay between verification and testing.
Another idea is to employ a verification-driven development approach, similar to test-driven development~\cite{tdd}, where for each piece of functionality, the specification is formalized (and potentially even validated) first and the code is written and iteratively refactored afterwards, aiming to satisfy the specification.

\conclusion[concl:ru2]{RU2}{Concentrate verification efforts on critical components}
Given the high cost of full formal verification, we recommend focusing on critical properties of critical components, such as security mechanisms, concurrency models, or safety logic. 
Separating and encapsulating these components can aid in reducing the interactions with unverified code and, thus, unverified assumptions.
For less critical parts, using lightweight methods such as testing or static analysis can ensure quality without the overhead of full proofs. 
This dual approach weakens formal guarantees to some extent, but helps keep verification effort manageable. 

\conclusion[concl:ru3]{RU3}{Invest in training and education}
Since formal verification requires specialized expertise, investment in training for engineers is crucial to promote adoption and long-term success.
Structured learning opportunities, such as hands-on practice with verification tools, aid in flattening the steep learning curve.
Naturally, the effectiveness of these initiatives is predicated on the availability of extensive learning resources provided by experts (see also \nameref{concl:rd6}).

Moreover, given the current scarceness of relevant expertise, we recommend prospective users seek collaborations with developers of tools they are considering using.
Various of our interviewees reported on finding this type of support highly helpful.
Involving experts and tool developers helps counter the ``black box'' nature of many verifiers and also identify areas of improvements in tools.

\subsubsection{Recommendations for Tool Developers and Researchers}\label{sec:04-recommendations-devs}
To overcome the barriers outlined in Sec.~\ref{sec:04-barriers}, we recommend the following seven issues be addressed.

\conclusion[concl:rd1]{RD1}{Develop methods integrating verification with other quality assurance techniques}
Beyond merely focusing on deductive verification alone, future research could explore how different quality assurance methods could more tightly integrate, compensating for each other's weaknesses.
For instance, bridging the divide between model checking and deductive verification could help in verifying liveness properties or concurrent code more approachable.
Failed tests after refactoring could inform proof repair and thus aid in proof maintenance.

Providing strong guarantees in the presence of unverified code is an important research challenge. 
Runtime monitoring could detect executions violating assumptions made by verified parts.
Similarly, such assumptions might presuppose the absence of issues like null pointer dereferences, which static analysis can detect without program execution.
Finally, assumptions introduced in verified components could guide test generation for unverified parts, documenting dependencies, revealing assumption violations, and strengthening the overall guarantees.

\conclusion[concl:rd2]{RD2}{Facilitate the management of specifications and annotations}
Automatically generating or inferring the annotations required by the verifier could significantly increase efficiency and lower the entry barrier to using verification tools.
In addition to classical methods such as abstract interpretation and abduction, AI-driven techniques could assist users by suggesting annotations, filling in missing steps, or even adapting proofs to different contexts and assisting in proof maintenance.
Such techniques could potentially also generate formal specifications from high-level requirements.

Domain-specific languages (DSLs) for formal requirements provide intuitive, high-level abstractions aligned with specific domains such as law~\cite{catala}.
As formalizing requirements is a challenge in itself, especially for non-experts, such DSLs potentially facilitate this process by reducing complexity and enabling more natural expression of the desired properties.

\conclusion[concl:rd3]{RD3}{Invest in tool usability}
Our data suggests that usability plays a critical role for verification.
Most notably, alleviating the burden of manual proof engineering by employing automation will make it easier for developers, especially those with limited formal verification experience, to incorporate verification into their daily practices.
Automation can be helpful in any stage of the verification process:
It could be employed in early stages, for instance, for writing or validating specifications, and later, e.g., for generating proof annotations or adjusting annotations and specifications to changes in the code.
Complementing automated solutions with features supporting interactiveness ensures that tools can scale to complex projects and meet the needs of experienced professionals as well (see also \nameref{concl:rd4}).

In general, verification frameworks would benefit from powerful IDEs and elaborate tooling.
Beyond standard features such as syntax highlighting, autocompletion, or run configurations, a particularly important aspect is support for debugging verification errors, which current frameworks do not yet address satisfactorily.

\conclusion[concl:rd4]{RD4}{Balance automation with interactiveness}
As verification projects become more ambitious, complementing automated verification frameworks with support for user intervention becomes an important avenue for research:
On the one hand, it is desirable to design tools to have a low entry barrier, i.e., to be easy to use, especially for engineers without an extensive background in verification.
On the other hand, tools should be sophisticated and expressive enough to provide dedicated support for more advanced users who require detailed information about and more fine-grained control of the verification process.
Following this low-entry-high-ceiling approach, i.e., striking a balance between automation and user interaction, is vital for making verification tools both powerful and accessible, yet state-of-the-art solutions still fall short of achieving this balance.

Engineers should be able to comprehend how verification tools operate, e.g., through detailed logs, interactive visualizations of proof obligations, memory layouts, or quantifier instantiations, access to intermediate states, and other insights that make the verification process more transparent.
Additionally, better support for guiding verifiers -- such as interactive proof exploration or configurable automation levels -- can enable users to steer the verification process more effectively.
Such features facilitate debugging and build trust in automated reasoning.

\conclusion[concl:rd5]{RD5}{Improve the performance of automated tools}
Performance was identified as a critical hurdle in SMT-based verification. 
Besides optimizing tools directly, it is promising to explore support for currently manual performance optimizations, such as proof decomposition, which existing tools do not yet adequately provide.
Tools could assist the user in deciding where and how to split up proofs, e.g., by identifying reusable reasoning patterns or pinpointing complex proof obligations.

Breaking down verification into more tractable components and later combining them to form a full proof improves both the speed and feasibility of the verification process.
Modularization could be achieved along various dimensions, for instance, by program components (e.g., modules, functions, or threads), by properties (e.g., safety, functional correctness, or security), or by refinement steps. 
Tools should support all of these and allow for the seamless combination of subproofs.

\conclusion[concl:rd6]{RD6}{Facilitate proof maintenance}
Maintaining proofs as software evolves is a major challenge. 
Since code changes are inevitable, addressing this open issue is essential.

First, code changes may break existing proofs, typically because one needs to adjust existing annotations such as loop invariants. 
Thus, it is important to develop effective proof repair techniques. 
In comparison with general inference, these techniques can potentially be more effective because they can utilize the proof for the prior version of the code.

Second, verified components that interact with unverified code need to make assumptions about the latter, which are typically validated informally through reviews and testing, as became evident in our interviews. 
Changes to the unverified code potentially violate these assumptions, thereby invalidating the existing proofs. 
Since verification tools cannot detect such situations, it is vital to develop other quality assurance techniques to validate assumptions efficiently.

\conclusion[concl:rd7]{RD7}{Investigate the prevalent approach to teaching verification}
Besides technical aspects of verification, our interviews highlight many \emph{human} hurdles for adoption, ones not traditionally addressed by research. 
Verification is described as requiring a different way of thinking, having a steep learning curve, and being frustrating to decipher failures.

To address the first point, the community would benefit from reevaluating the prevalent approach to pedagogy: 
Courses on verification in universities that currently place a strong emphasis on theoretical foundations should also include lectures dedicated to the practical application of state-of-the-art tools, putting the theory into practice. 
Additionally, or alternatively, one could design entire courses on applied formal methods or tool-supported verification, where students undertake practical case studies focused on specifying and verifying complete systems, rather than working through small, isolated exercises -- similar to the structure of software engineering courses. 
Such initiatives would allow students to gain hands-on experience, equipping them with the necessary skills to apply verification in large-scale projects.

Second, to reduce the complexity of learning verification tools, developers should provide comprehensive tutorials on how to use their framework~\cite{coq-tutorial,isabelle-tutorial,lean-tutorial,dafny-tutorial,viper-tutorial,fstar-tutorial,verus-tutorial}.
To ensure feasibility for industrial contexts, these self-study programs should not (only) focus on teaching students, but target industry professionals.

Lastly, the verification community has inherited the functional programming community's practice of \emph{pearls}~\cite{verifythis2021,proofpearl1,proofpearl2,proofpearl3}, highlighting elegant proofs of small programs.
Such proofs could form the basis for a systematic catalog of ``idiomatic'' proofs of programs, helping intermediate users navigate the plethora of choices to be made when starting verification. 

\section{Related Work}\label{sec:05}
Numerous experience reports and case studies elaborate on individual projects using deductive verification, for instance, in cloud computing, compiler design, Internet computing, and other areas~\cite{aws,sel4,cakeml,compcert,verified-scion,ironfleet,ironclad}. Here, we focus on surveys and empirical studies beyond individual case studies. Due to significant recent advancements, we limit related work to the past ten years.

To the best of our knowledge, our study is the first to present an in-depth discussion of the practical usage of \emph{deductive} verification, grounded in data collected across the entire community. 
Differences in the characteristics of verification techniques, which have substantial implications for their practical application, warrant a focused study.

Nyberg et al.~\cite{nyberg} summarize six case studies, one of which on deductive verification~\cite{gurov}, and discuss enablers and barriers when employing formal methods in the automotive industry.
While our findings align with some of their conclusions, e.g., the importance of automation, our work provides a broader discussion of obstacles involving all phases in the development process, such as maintenance, which Nyberg et al.\ do not explicitly consider. 
The use of formal methods, among others deductive verification tools, in the railways industry was investigated by ter Beek et al.~\cite{terbeek1,terbeek2} via questionnaires. 
A report by Beckert et al.~\cite{beckert} focuses on the deductive verification of legacy code, drawing on experience with two case studies. 
Notably, neither ter Beek et al.\ nor Beckert et al.\ address tool performance, an issue that was identified in our study as a major barrier.
All of these publications are limited in their scope to case studies or application areas, whereas we conducted interviews across several domains, making our results more generalizable and nuanced.

In regards to formal methods in general, recent work by ter Beek et al.~\cite{terbeek3} presents a detailed summary of concrete applications of formal methods in various domains and concludes with advocacy and ideas for integrating their use better into computer science education.
Reid et al.~\cite{reid} published a paper on how to unlock wider adoption of formal methods, basing their proposals on personal experience and existing literature.
Similarly, a 2020 essay by Huisman et al.~\cite{huisman} explores the gap between the capabilities of formal methods and their employment in industry and presents high-level, experience-based recommendations to address accessibility, focusing on human factors rather than open research problems.
Our study confirms several of their intuitions by employing the scientific method, e.g., the importance of smooth integration into existing development workflows, and discusses a plethora of aspects not considered in their work, such as the role of automation or maintenance concerns.
Importantly, we concretize Huisman et al.'s results by providing guidelines for practitioners and research opportunities.

Empirical studies on formal methods include a paper by Garavel et al.~\cite{garavel} presenting a survey about the state of formal methods in 2020 among experts in the field. 
The high-level conclusions of the survey align with our results, particularly regarding key obstacles in the adoption of formal methods, such as the lack of expertise, tool usability, and performance.
Our study offers a more detailed view of these concerns by identifying specific tool features practitioners value or desire, for example, the ability to verify isolated code fragments for better scalability, support for intuitive counterexamples and axiom profiling, and tooling improvements like support for debugging verification errors.
More importantly, though, we raise issues that Garavel et al.\ do not explicitly discuss, for instance, the insufficiently addressed challenges related to proof maintenance, and we go beyond data collection and analysis by synthesizing concrete recommendations for various stakeholders, derived from our interview data.

The use of model checking has also been addressed in the literature~\cite{kurshan,ovatman}, as well as static analysis~\cite{christakis}.
A notable parallel between model checking and our work is the challenge that tool performance poses. 
Tool performance and usability were also two of the main concerns in the study on static analysis.
In the realm of testing, Goldstein et al.~\cite{pbt} investigated property-based testing; their work inspired our study.

\section{Conclusion}\label{sec:06}
We presented the first  study of the practical usage of deductive verification, grounded in data collected across the entire community.
Beyond empirically confirming folklore wisdom, our work surfaced several major challenges underrepresented in the scientific discourse, such as proof maintenance, insufficient support for user intervention in automated tools, lack of usability, and problems concerning the integration of verification.
Based on these findings, we synthesize future research and development directions that are grounded in empirical evidence, such as developing support for (manual) mitigation strategies for performance limitations, designing tools that offer both automation and control, and devising verification techniques that give strong guarantees in the presence of unverified code.

Follow-up work on this paper could explore the discussed topics in more detail; for example, a qualitative investigation of tool usability or accessibility would shed more light on the concrete challenges (new) users are faced with.

\section*{Data Availability}\label{artifact}
Due to confidentiality and the need to protect participant anonymity, the interview transcripts serving as a basis of our data analysis cannot be publicly shared.
However, the full codebook used for coding the interviews, along with the aggregated results of the pre-interview questionnaire, is available online~\cite{artifact}.

\bibliographystyle{ACM-Reference-Format}
\bibliography{bibliography}

\end{document}